\documentclass[11pt]{article} 

\usepackage{mathtools, amsfonts, amsthm, latexsym, amssymb,dsfont}
\usepackage[T1]{fontenc}
\usepackage[tracking=true]{microtype}
\usepackage{lmodern}
\usepackage{stmaryrd}
\usepackage{IEEEtrantools}
\usepackage{float}

\usepackage[margin=1in]{geometry}

\usepackage{stmaryrd}
\usepackage{amsmath}
\usepackage{amsfonts}
\usepackage{amssymb}
\usepackage{physics}
\usepackage{braket}
\usepackage{tensor}
\usepackage{simplewick}

\usepackage{enumitem}
\usepackage{graphicx}
\usepackage{xcolor}
\usepackage[utf8]{inputenc}
\usepackage{caption,subfig}
\usepackage{bm,bbold}	
\usepackage{tikz}	
\usepackage{scalerel}
\usepackage{feynmp-auto}
\usepackage{booktabs}
\usepackage{longtable}
\usepackage{verbatim}
\usepackage{cite}

\usepackage{color}
\definecolor{darkblue}{cmyk}{0.9,0.9,0,0}
\definecolor{wine-stain}{rgb}{0.5,0,0}
\usepackage[colorlinks, allcolors=darkblue, linktoc=all]{hyperref} 
\usepackage{here}

\newcommand{\E}{}
\def\E#1#{\tensor#1{\epsilon}}

\renewcommand{\L}{\mathcal L}

\newcommand{\pd}{\partial}
\newcommand{\mr}[1]{\mathrm{#1}}
\def \be {\begin{equation}}
\def \ee {\end{equation}}

\begin{document}
\thispagestyle{empty}
\vspace*{1in} 

\begin{center}
{\Large \textbf{Quantum Stability of Generalized Proca Theories}}
\vspace{.6in}

\textbf{Lavinia Heisenberg and Jann Zosso}
\vspace{.2in} 

{\small \textit{Institute for Theoretical Physics, ETH Zurich, Wolfgang-Pauli-Strasse 27, 8093, Zurich, Switzerland}}
\end{center}
\vspace{.2in}

\begin{abstract}
We establish radiative stability of generalized Proca effective field theories. While standard powercounting arguments would conclude otherwise, we find non-trivial cancellations of leading order corrections by explicit computation of divergent one-loop diagrams up to four-point. These results are crosschecked against an effective action based generalized Schwinger-DeWitt method. Further, the cancellations are understood as coming from the specific structure of the theory through a decoupling limit analysis which at the same time allows for an extension of the results to higher orders.
\end{abstract}

\setcounter{page}{0}

\newpage


\section{Introduction}\label{Sec:Introduction}

Over the last century the theory of General Relativity accumulated a rock solid empirical foundation on a broad band of scales with tests ranging from high precision laboratory experiments to the observation of the predicted gravitational waves traveling through the fabric of space and time \cite{Will:2005va,Abbott:2016blz}. However, almost from beginning the beauty of the theory was smudged by the apparent absence of gravitating vacuum energy, the so called cosmological constant problem \cite{RevModPhys.61.1,Martin:2012bt}, a strong indication that Einsteins theory might not be the end of the story on IR gravity.
On top of this, the evidence for the current accelerating expansion of the universe \cite{Perlmutter:1998np,Riess:1998cb} additionally drives the search for a plausible generalization of the theory of gravity on cosmological scales with the hope that dynamical dark energy could perhaps at the same time provide a mechanism which screens the cosmological constant. 
 
There exist a multitude of ideas for consistent extensions of GR \cite{Clifton:2011jh,Copeland:2006wr,Heisenberg:2018vsk}. As it is the unique EFT of a massless spin 2 degree of freedom in four dimensions\footnote{Up to reasonable assumptions.}, extending it almost inevitably introduces additional degrees of freedom. In a field theory framework, the new degrees of freedom typically manifest themselves as additional scalar, vector or tensor fields. While it appears as a rather easy task to just throw in new degrees of freedom in order to modify gravity on large scales, the challenge is to simultaneously do justice to the unquestioned success of GR on smaller scales and denser regions. Hence, the newly introduced fields must effectively decouple from matter in these ranges.

In that respect, theories which contain higher order derivative self-interactions become interesting, as they naturally incorporate a Vainshtein screening mechanism \cite{VAINSHTEIN,Deffayet:2001uk,Kimura:2011dc,Babichev:2013usa,Koyama:2013paa,Kase:2013uja}. This mechanism essentially relies on the non-linearities becoming large near a massive source, such that the kinetic term of perturbations gets enhanced significantly, which in turn weakens their interaction with matter. 

In general, theories with derivative self-interactions suffer from Ostrogradsky instabilities \cite{Ostrogradsky:1850fid,Woodard:2015zca}, propagating a ghost degree of freedom. However, in certain cases it is possible to construct theories which evade this rule. A prominent example are the scalar Galileon theories in flat spacetime \cite{Nicolis:2008in}, whose finite amount of non-linear derivative interaction terms are composed in such a way that they nevertheless lead to second order equations of motion and thus still only propagate the desired degree of freedom. Asking theoretical consistency, this immediately leads to the question whether these classical interactions are stable under quantum corrections. Naively, as the Vainshtein mechanism relies on scales for which non-linear interactions are large compared to the kinetic term, one could expect that the EFT is not protected against equally non-renormalizable quantum corrections. At a closer look, however, the EFT is organized in such a way that there exist a regime for which classical non-linearities dominate, while quantum effects are still under control \cite{Luty:2003vm,Nicolis:2004qq,Burgess:2006bm,Hinterbichler:2010xn,Goon:2016ihr}: All terms generated by quantum loops have more derivatives per fields compared to the nonlinear galileon interactions.\footnote{This is in close analogy to the EFT structure of GR: Diffeomorphism invariance protects the relative coefficients of the classical non-linear terms from detuning. Below the plank scale, other terms generated through loop contributions are suppressed, even in regimes where the classical non-linearities responsible for all high curvature effects become important. Note, however, that the cosmological constant problem spoils the perfect IR picture.} This provides the EFT Lagrangian with two distinct expansion parameters, which allow for regions below the UV cutoff scale, notably dense regions with non-negligible curvature, where classical non-linearities become important and the Vainshtein mechanism screens the coupling of the scalar field to matter, while quantum corrections are still under control. On large scales, both classical and quantum derivative self interactions become negligible, such that the scalar degree of freedom can be used as an extension of classical gravity. 

Various counterterms of the galileon EFT have been calculated explicitly \cite{dePaulaNetto:2012hm,Brouzakis:2013lla,Heisenberg:2019udf,Heisenberg:2019wjv} and the theory has been generalized to arbitrary spacetimes \cite{Deffayet:2009wt,Deffayet:2009mn} which lead to a rediscovery of the most general scalar-tensor theory with second order equations of motion \cite{Horndeski:1974wa}. These Horndeski theories and associated generalizations have found various applications in cosmology \cite{Deffayet:2010qz,Appleby:2011aa,Deffayet:2011gz,Kobayashi:2011nu,Appleby:2012ba,Barreira:2012kk,Okada:2012mn,Bartolo:2013ws,Creminelli:2012my,Neveu:2013mfa,Barreira:2013jma,Barreira:2013eea,Gleyzes:2014dya,HorndeskiSurvivals}, in particular, galileon theories naturally arise as the zero-helicity part of the graviton in higher dimensional models \cite{Dvali:2000hr} and massive gravity theories (see \cite{Hinterbichler:2011tt,deRham:2014zqa} for reviews).

In a cosmological context, scalar fields are by far the most popular choice when it comes to adding new degrees of freedom, as they naturally go along with the basic assumptions of homogeneity and isotropy. At the same time, this means that throwing in any desired amount of new scalar dofs is very cheap, in the sense that there are a priori not many restrictions on how to introduce them and the space of possibilities seems endless. It could very well be, that todays inconsistencies in the theory of cosmology require a light departure from the convenient simplifying basic assumptions. This should serve as a motivation to consider the other possibilities at hand. 

For instance, when endowing an abelian spin one field with a mass, it's temporal component can readily serve as an isotropic starting point, with non-abelian cases allowing for even richer structures. Interestingly, a massive vector field\footnote{In contrast to the gauge symmetric case, where a no-go theorem for consistent derivative self-interactions has been proven \cite{Deffayet:2013tca}.} also admits a galileon-like ghost free structure of higher order derivative interactions usually referred as generalized Proca theory \cite{Heisenberg:2014rta,Allys:2015sht,Jimenez:2016isa}, which inherits the benefits of a naturally incorporated Vainshtein screening \cite{DeFelice:2016cri}. Indeed, Proca theories and their various generalizations have already appeared in a cosmological context on various promising occasions \cite{Boehmer:2007qa,Golovnev:2008cf,Jimenez:2013qsa,BeltranJimenez:2013fca,Tasinato:2013oja,Hull:2014bga,Khosravi:2014mua,Tasinato:2014eka,Hull:2015uwa,Jimenez:2015fva,Jimenez:2016opp,Heisenberg:2016eld,Kimura:2016rzw,Heisenberg:2016lux,Jimenez:2016upj,Allys:2016kbq,Lagos:2016wyv,DeFelice:2016yws,DeFelice:2016uil,Heisenberg:2016wtr,Emami:2016ldl,Rodriguez:2017wkg,deFelice:2017paw,Heisenberg:2018acv,Petrov:2018xtx,ErrastiDiez:2019trb}. 

The generalized Proca theories, also known as vector Galileons, possess an intimate relation to scalar Galileons. At high energies way above the vector mass the longitudinal polarization dominates, such that the theory acquires a Galileon symmetry and half of the generalized Proca interactions reduce to pure scalar Galileon terms. In particular, in parallel to it's scalar counterpart the organization of the generalized Proca EFT is highly non-trivial. A crucial step in the analysis of the theoretical viability of any EFT is it's quantum stability. Yet, a thorough analysis of the behavior of generalized Proca theories under loop corrections is in large parts still missing and filling this gap is the goal of the present work. The absence of a particular symmetry of the interactions makes it unlikely that the classical structure is protected from quantum detuning, as also indicated by an earlier result \cite{Charmchi:2015ggf}. Nevertheless, we claim radiative stability of generalized Proca EFT's in the sense that quantum loop corrections remain suppressed enough that the mass of henceforth introduced ghost degrees of freedom reside safely above the EFT cutoff.

In \S\ref{GPm} we first introduce the particular generalized Proca model we chose for the analysis and reformulate the theory by introducing a scalar St\"uckelberg field. Section \ref{Feyn} is then devoted to the explicit calculation of one-loop UV divergences of Feynman diagrams up to four external legs. In doing so, we correct results obtained in \cite{Charmchi:2015ggf} and generalize the analysis to a more complete picture. These results are consolidated by means of an effective action based generalized Schwinger-DeWitt method in \S\ref{Schw}. Decoupling limit arguments in \S\ref{QuantumStability} then allow for an interpretation of the obtained results and enable us to go one step further by finding strong indications for quantum stability of the vector Galileon theory in its full generality, which means including all possible generalized Proca operators and an extension to higher loop orders.


\section{Generalized Proca Model and St\"uckelberg Formulation}
\label{GPm}

The most general Lagrangian of a local massive vector field theory with second order equations of motion and three propagating degrees of freedom is restricted to the following structure \cite{Heisenberg:2014rta,Jimenez:2016isa}:

\small
\begin{IEEEeqnarray}{rCl}\label{Lagrangians}
\L_2 &=&  \Lambda_2^4\, f_2 \left(\scriptstyle\frac{mA_\mu}{\Lambda_2^2},\frac{F_{\mu\nu}}{\Lambda_2^2},\frac{\tilde{F}_{\mu\nu}}{\Lambda_2^2}\right)\, ,\nonumber \\
 \L_3 &=& - \, \frac{\Lambda_2^2}{6} \, f_3 \left({\scriptstyle \frac{m^2A^2}{\Lambda_2^4}}\right) \, \E{^\mu^\nu^\rho^\sigma}\E{^\alpha_\nu_\rho_\sigma} \, \pd_\mu A_\alpha\,, \nonumber\\
\L_4 &=& -  \,\frac{1}{2}\;\E{^\mu^\nu^\rho^\sigma}\E{^\alpha^\beta_\rho_\sigma} \left(\, f_4 \left({\scriptstyle  \frac{m^2A^2}{\Lambda_2^4}}\right)\pd_\mu A_\alpha \, \pd_\nu A_\beta + \tilde{f}_4 \left({\scriptstyle  \frac{m^2A^2}{\Lambda_2^4}}\right) \pd_\mu A_\nu \, \pd_\alpha A_\beta \right), \\
\L_5 &=& -\frac{1}{\Lambda_2^2}\, \E{^\mu^\nu^\rho^\sigma}\E{^\alpha^\beta^\gamma_\sigma} \left(f_5 \left({\scriptstyle  \frac{m^2A^2}{\Lambda_2^4}}\right)\pd_\mu A_\alpha \, \pd_\nu A_\beta \,\pd_\rho A_\gamma + \tilde{f}_5 \left({\scriptstyle  \frac{m^2A^2}{\Lambda_2^4}}\right) \pd_\mu A_\nu\, \pd_\alpha A_\beta \,\pd_\rho A_\gamma \right),\nonumber \\
\L_6 &=& - \frac{1}{\Lambda_2^4} \, \E{^\mu^\nu^\rho^\sigma}\E{^\alpha^\beta^\gamma^\delta} \left(f_6 \left({\scriptstyle  \frac{m^2A^2}{\Lambda_2^4}}\right)\pd_\mu A_\alpha\, \pd_\nu A_\beta\, \pd_\rho A_\gamma\, \pd_\sigma A_\delta + \tilde{f}_6 \left({\scriptstyle  \frac{m^2A^2}{\Lambda_2^4}}\right) \pd_\mu A_\nu\, \pd_\alpha A_\beta\, \pd_\rho A_\gamma\, \pd_\sigma A_\delta \right),\nonumber
\end{IEEEeqnarray}
\normalsize
where the two classical scales of the theory are the mass $m$ and the interaction scale $\Lambda_2$ which controls the interactions expanded in the number of fields $n$ through factors of \small$\frac{1}{\Lambda_2^{2n-4}}$\normalsize. The dimensionless combination $m/\Lambda_2$ can be viewed in some sense as a coupling constant, generally assumed to be small. The numerical factors in the definitions of $\L_3$ and $\L_4$ are pure convenience.

The lagrangian term $\L_2$ contains all possible potential contributions including the mass term, as well as kinetic and interaction terms constructed out of the building blocks $A_\mu$, it's field strength $F_{\mu\nu}$ and the dual $\tilde{F}^{\mu\nu}\equiv\frac{1}{2}\E{^\mu^\nu^\rho^\sigma}F_{\rho\sigma}$, which by construction do not give rise to any dynamics of the temporal component $A_0$. On the other hand, $\L_{3,.., 6}$ represent derivative self-interactions which nevertheless remain ghost-free and thus only propagate the required three degrees of freedom \cite{Heisenberg:2014rta}. This is ensured by their construction via two Levi-Civita tensors which at the level of the equations of motion only allows for at most second order terms restricted to the very specific gauge invariant form $\pd F$.\footnote{This follows from the fact that two derivatives applied on the same field can only enter through $
\E{^\mu^\nu^\rho^\sigma}\E{^\alpha^\beta_\rho_\sigma}\pd_\mu\pd_\alpha A_\beta\sim\pd_\mu\pd^\mu A^\nu-\pd_\mu\pd^\nu A^\mu=\pd_\mu F^{\mu\nu}$, while all other derivative terms remain first order.}

Being interested in quantum corrections which potentially renormalize the given classical structure we will choose a minimal model with standard canonically normalized kinetic and mass term and where
\be
\label{FunctionChoice}
 f_{3,4}(x)=c_{3,4}\,x \; ,\quad \tilde{f}_4(x)=\tilde{c}_{4}\,x  \; ,\quad f_{5,6}(x)=c_{5,6} \; ,\quad \tilde{f}_{5,6}(x)=\tilde{c}_{5,6}\, .
\ee
With this choice the terms proportional to $c_5$ and $c_6$ are total derivatives and effectively drop out of the analysis, while the other terms up to total derivatives take on the form\footnote{Throughout this work we will employ a mostly minus metric-sign convention $(+,-,-,-)$.}
\begin{equation}\label{action}
\begin{split} 
& \L_{2}=-\frac{1}{4}F^2+\frac{1}{2}m^2A^2\,,\\
& \L_3 = \frac{m^2}{\Lambda_2^2} \,c_3\, A^2 \pd\cdot A\,, \\
& \L_4 = \frac{m^2}{\Lambda_2^4}\,A^2\,\left(c_4\left[(\pd\cdot A)^2-\pd_\mu A_\nu\pd^\nu A^\mu\right]+\tilde{c}_4\,F^2\right)\,, \\
& \L_5 =  -\frac{1}{\Lambda_2^2}\,\tilde{c}_5\, \E{^\mu^\nu^\rho^\sigma}\E{^\alpha^\beta^\gamma_\sigma} \pd_\mu A_\nu\, \pd_\alpha A_\beta \,\pd_\rho A_\gamma \, \\
&  \L_6 = - \frac{1}{\Lambda_2^4}\,\tilde{c}_6\,\E{^\mu^\nu^\rho^\sigma}\E{^\alpha^\beta^\gamma^\delta} \pd_\mu A_\nu\, \pd_\alpha A_\beta\, \pd_\rho A_\gamma\, \pd_\sigma A_\delta \,.
\end{split}
\end{equation}
Note that the operator proportional to $\tilde{c}_4$ is actually a higher order $\L_{2}$ term. We will nevertheless keep it in order to explicitly see what happens with this class of terms.
The vector propagator of the theory reads
\be\label{propagator}
D_{\mu\nu}(x-y)=\int \frac{\mr{d}^4p}{(2\pi)^4}\,\mr{e}^{ip(x-y)}\; i\, \frac{-\eta_{\mu\nu}+{\scriptstyle \frac{p_\mu p_\nu}{m^2}}}{p^2-m^2}
\ee
with implicit Feynman-prescription.

It will be useful in the following to rewrite this theory of a self-interacting massive vector field by introducing a redundancy in the form of an additional scalar field $\phi$ through the replacement\footnote{It is important to note that this replacement is not a change of field variables and neither a decomposition of $A_\mu$ into transverse and longitudinal degrees of freedom. It merely introduces redundancy in the description.}
\be\label{Stuckelberg}
A_\mu\rightarrow A_\mu+\tfrac{1}{m}\partial_\mu\phi \,.
\ee
where the mass scale is fixed by canonically normalizing the kinetic term of the scalar field. This formulation goes back to the work of St\"uckelberg \cite{Stueckelberg:1900zz,Ruegg:2003ps} and can be viewed as an explicit reintroduction of the eaten Goldstone boson. The specific form of the replacement \eqref{Stuckelberg} is such that gauge invariant terms remain untouched and moreover suggests the definition of a covariant derivative $D_\mu\phi\equiv\pd_\mu\phi+m A_\mu$. The new theory is thus effectively obtained by making the replacements
\be\label{Stuckelberg2}
A_\mu\rightarrow \frac{1}{m}D_\mu\phi \,,\quad F\rightarrow F \quad\text{and} \quad\tilde{F}\rightarrow \tilde{F}
\ee
in \eqref{action}. This renders the theory invariant under the simultaneous gauge transformation
\be\label{gaugeS}
\phi\rightarrow\phi+m\,\alpha\, ,\quad A_\mu\rightarrow A_\mu-\pd_\mu\alpha\,.
\ee
Note that the unitary gauge choice $\alpha=-\frac{\phi}{m}$ sets $\phi=0$, which shows that the new theory is indeed equivalent to \eqref{action} and only propagates three degrees of freedom. Through a different gauge choice $\pd_\mu A^\mu+m\phi=0$ implemented in a Fadeev-Popov procedure one obtains the propagators of $A_\mu$ and $\phi$ \cite{Hinterbichler:2011tt}
\be
\frac{-i\,\eta_{\mu\nu}}{p^2+m^2}\quad\text{and}\quad \frac{-i}{p^2+m^2}\,,
\ee
which at high energies behave as $\sim \frac{1}{p^2}$ compared to $\sim\frac{1}{m^2}$ in the old formulation \eqref{propagator}.

The lowest strong coupling scale of the theory is found by looking at the pure scalar sector. For instance, the $2\rightarrow 2$ tree-leel amplitude coming from the operator of the schematic form $\sim\frac{m^2}{\Lambda_2^4}\frac{1}{m^4}(\pd\phi)^2(\pd^2\phi)^2$ in $\L_4$ goes like $\mathcal{M}_{\scriptstyle2\rightarrow 2}\sim\frac{E^6}{\Lambda_2^4m^2}$, such that at energies above the scale
\be
\Lambda_3\equiv \left(\Lambda_2^2m\right)^{\frac{1}{3}}\,,
\ee
 the theory becomes strongly interacting. Note that as long as $m^2\ll \Lambda_2$ (small classical coupling constant) this new scale is separated from the vector mass $m$ by a parametrically large gap, which is essential for the healthiness of the EFT. Moreover, it is a requirement for the decoupling limit to be valid. In this limit, one zooms into the cutoff $\Lambda_3$ of the theory by keeping it fixed, while sending the smaller and higher scales away to zero and infinity respectively
\be\label{DL}
m\rightarrow 0 \quad\text{and}\quad \Lambda_2\rightarrow \infty\,,\;\;\text{while}\quad \Lambda_3\equiv\left(\Lambda_2^2m\right)^{\frac{1}{3}}=\text{const.}
\ee
This decouples in large parts the vector from the Goldstone boson, as the vector field only survives in the gauge invariant combinations $F$ and $\tilde{F}$. In particular, the coupled gauge symmetry \eqref{gaugeS} is broken apart and only the one of $A_\mu$ prevails, while the scalar field merely retains an independent global shift symmetry
\be\label{gaugeDL}
\phi\rightarrow\phi+c\, ,\quad A_\mu\rightarrow A_\mu-\pd_\mu\alpha\,.
\ee
The surviving terms can directly be obtained from the original theory \eqref{action} by replacing
\be\label{StuckelbergDL}
A_\mu\rightarrow \frac{1}{m}\pd_\mu\phi \,,\quad F\rightarrow F \quad\text{and} \quad\tilde{F}\rightarrow \tilde{F}\,,
\ee
and applying the limit \eqref{DL} which yields
\begin{equation}\label{actionDL}
\begin{split} 
& \L_{2}=-\frac{1}{4}F^2+\frac{1}{2}(\pd\phi)^2\,,\\
& \L_3 = \frac{1}{\Lambda_3^3} \,c_3\, (\pd\phi)^2 \Box \phi\,, \\
& \L_4 = \frac{1}{\Lambda_3^6}\,(\pd\phi)^2\,c_4\,\left[(\Box \phi)^2-\left(\pd_\mu \pd_\nu\phi\right)^2\right]\,, \\
& \L_5 = -\frac{1}{\Lambda_3^3}\,\tilde{c}_5\,\tilde{F}^{\mu\alpha}\tilde{F}^{\nu}_{\;\alpha} \pd_\mu \pd_\nu\phi\,, \\
&  \L_6 = -\frac{1}{\Lambda_3^6}\,\tilde{c}_6\,\tilde{F}^{\mu\alpha}\tilde{F}^{\nu\beta} \pd_\mu \pd_\nu\phi\pd_\alpha \pd_\beta\phi\,.
\end{split}
\end{equation}
Hence, the mass term of the vector field and the term proportional to $\tilde{c}_4$ vanish, while the terms proportional to $c_3$ and $c_4$ reduce to pure scalar Galileon interactions \cite{Nicolis:2008in}.

From \eqref{StuckelbergDL} it follows that the decoupling limit can alternatively be viewed as a high energy limit in the original theory, in the sense that scaling down $m$ is the same as scaling up the energy in the factor $\frac{\partial}{m}$. This makes contact with the Goldstone boson equivalence theorem: At rest, all three polarizations are equivalent, but at higher energies, the transverse polarizations and the rapidly moving longitudinal polarization are clearly distinguished. The decoupling limit is thus particularly useful when analyzing the quantum stability of an EFT, as it focuses on the high energy behavior right at the relevant scale, while ignoring all others. Moreover, in this limit $A_\mu$ exclusively propagates the transverse modes with the gauge symmetry \eqref{gaugeDL} ensuring the absence of ghost instabilities and quantum detuning. In the decoupling limit, the analysis of the radiative stability of the EFT is thus reduced to an analysis of the behavior of the Goldstone. However, before being able to perform a thorough hierarchy classification of terms in the EFT in \S\ref{QuantumStability}, an explicit calculation of the most important counterterms at one-loop in section \ref{OneLoop} is in order.


\section{One-loop Corrections}
\label{OneLoop}

For now we will stick to the original formulation of the theory \eqref{action} which corresponds to a unitary gauge choice and analyze it's radiative stability. After discussing a general powercounting, explicit calculations of the logarithmic divergent part of the $1$PI Feynman diagrams up to four external legs are presented, crosschecked with a perturbative calculation based on the generalized Schwinger-DeWitt method developed in \cite{BARVINSKY19851}. In doing so, we for instance correct and extend the previous work of Charmchi \textit{et al.} \cite{Charmchi:2015ggf}. 

Because of the absence of any symmetry of the theory in the original formulation \eqref{action}, we do not expect the classical operators to be protected against quantum detuning. However, the corresponding quantum induced operators are always suppressed enough that the mass of the henceforth introduced ghost lies safely above the cutoff of the theory, as will be shown below using dimensional arguments. In dimensional regularization, the dangerous terms are really the operators at higher orders in derivatives which potentially lead to destabilizing ghost degrees of freedom. This is because the aforementioned modified high energy behavior of the vector propagator \eqref{propagator} allows the mass scale of the theory to appear in the denominator of the expressions and this enhancement can be enough that at energies close to the cutoff these quantum terms would dominate over the classical structure and thus destabilize the EFT description. This can be made explicit by taking the decoupling limit, in which these terms would then diverge. And indeed, a powercounting of the expected results (see eg. \eqref{2ptDim} below) indicates that such dangerous terms can very well arise in the theory. We will however explicitly show here that for all two- and three-point one-loop corrections, these dangerous leading order terms either cancel completely or take on an explicit gauge invariant form which effectively tames their bad high energy behavior. Hence, the corresponding induced operators do fit into the crucial EFT hierarchy between classical and quantum terms. In the next section we will then be able to show that these cancellations can generally be expected by analyzing the quantum behavior of the theory directly in the decoupling limit. This will set the ground for a complete analysis of the quantum stability of the generalized Proca theory offered in \S\ref{QuantumStability}.


\subsection{Feynman Diagram Calculation}
\label{Feyn}

In this section we compute the quantum behavior of the generalized Proca model in the unitary gauge at the one-loop level using standard Feynman diagram techniques. Each diagram represents a contribution to the reduced matrix element $\mathcal{M}_{1\text{PI}}$ in the perturbative expansion of the S-matrix:
\be\label{Smatrix}
\bra{k_{\text{out}}}\mathcal{S}\ket{k_{\text{in}}}\bigg\rvert_{1\text{PI}}=1+(2\pi)^4\,\delta^4(k_{\text{out}}-k_{\text{in}})\,i \mathcal{M}_{1\text{PI}} \;.
\ee
The reduced matrix element is calculated by summing over all possible Wick contractions of the form: 
\begin{equation}\label{Rules}
\begin{split} 
& \contraction{}{A}{_\mu(x)}{A}
A_\mu(x) A_\nu(y)=D_{\mu\nu}(x-y)\\
& \bcontraction{}{A^\mu}{(x)}{ket{k,\epsilon}}
A^\mu(x)\ket{k,\epsilon} = \epsilon^\mu_k\; \mr{e}^{-ikx} \\
& \bcontraction{}{ket{k,\epsilon}}{}{A}
\bra{k,\epsilon}A^\mu(x) = \epsilon^{*\mu}_k\; \mr{e}^{ikx} \, ,
\end{split}
\end{equation}
where the propagator $D_{\mu\nu}(x-y)$ was defined in \eqref{propagator} and $\epsilon^{\mu}_k$ denotes the associated polarization vector.

 Our minimal choices \eqref{FunctionChoice} only allow for vertices with up to four legs whose value depend on the derivative structure of the insertion that translates in fourier space into a dependence on the momenta which run on each leg. 

Following the $\overline{\text{MS}}$-scheme, the one-loop counterterms can then be inferred from the UV divergence of the $1$PI diagrams which we will extract using dimensional regularization. We are thus after the log-divergent part of the one-loop $1$PI diagrams with $N$ external legs $\mathcal{M}_N^{\text{div}}$ which will be a function of the external momenta $k_i$, ${\scriptstyle i=1,..,N-1}$, since the overall delta-function $\delta^4(k_{\text{out}}-k_{\text{in}})$ always allows to express one momentum $k_N$ in terms of the others. Throughout this work we will treat all momenta as incomming such that the overall delta-function translates to $\sum_{i=1}^N k_i=0$.

In the following we will calculate explicit divergent off-shell contributions up to four external legs and comment on their implications.

\subsubsection{Two-point}
\label{TwoPoint}

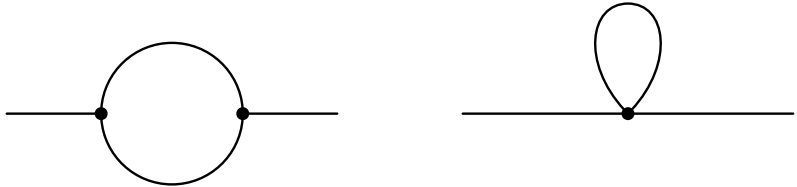
\begin{figure}[H]
\begin{center}
\begin{fmffile}{loops2pf}
\begin{fmfgraph*}(125,65)
     \fmfleft{i}
     \fmfright{o}
     \fmf{plain,tension=3}{i,v1}
     \fmf{plain,left=1}{v1,v2}
     \fmf{plain,left=1}{v2,v1}
     \fmf{plain,tension=3}{v2,o}
     \fmfdot{v1,v2}
    \end{fmfgraph*}
    \qquad \qquad
  \begin{fmfgraph*}(125,65)
 	\fmfleft{i}
     \fmfright{o}
     \fmf{plain,tension=3}{i,v1}
     \fmf{plain}{v1,v1}
     \fmf{plain,tension=3}{v1,o}
     \fmfdot{v1}
\end{fmfgraph*}
\end{fmffile}
\end{center}
\caption{\small{Two distinct one-loop $1$PI diagrams giving rise to corrections of the two point function. The first diagram represents contributions from the three possible combinations out of $\L_{3}$ and $\L_{5}$ and the second one separate contributions from $\L_{4}$ and $\L_{6}$. Each diagram comes with a symmetry factor of $2$.}}
\label{loopdia} 
\end{figure}

Within our minimal generalized Proca model characterized by the choices \eqref{FunctionChoice}, the perturbative renormalization procedure of the two-point function at one-loop requires the calculation of only two distinct $1$PI diagram structures depicted in Fig.\ref{loopdia} coming from $\L_{3,5}$ and $\L_{4,6}$ respectively.

Now, powercounting in dimensional regularization together with Lorentz invariance leads to the following expectation for the results
\small
\begin{equation}\label{2ptDim}
\begin{split} 
\mathcal{M}_{\scriptstyle\L_3\L_3}&\sim\,\frac{m^4}{\Lambda_2^4}\left(m^2+k^2+\frac{k^4}{m^2}+\frac{k^6}{m^4}\right)\,, \\
\mathcal{M}_{\scriptstyle\L_3\L_5,\,\L_4}&\sim\,\frac{m^2}{\Lambda_2^4}\left(m^4+m^2k^2+k^4+\frac{k^6}{m^2}+\frac{k^8}{m^4}\right)\,, \\
\mathcal{M}_{\scriptstyle\L_5\L_5,\,\L_6}&\sim\,\frac{1}{\Lambda_2^4}\left(m^6+m^4k^2+m^2k^4+k^6+\frac{k^8}{m^2}+\frac{k^{10}}{m^4}\right)\,,
\end{split}
\end{equation}
\normalsize
where $k$ stands for external momenta and the series presumably stops at $1/m^2$ or $1/m^4$, depending on how many propagators are involved in the loop. So indeed, classical structures are renormalized. For example from the terms proportional to $k^2$ we expect an explicit detuning of the gauge invariant kinetic combination by the introduction of a counterterm of the form
\be\label{Odetuning}
\sim\frac{m^4}{\Lambda_2^4}(\pd_\mu A^\mu)^2\,.
\ee
This term is however heavily suppressed such that the mass of the associated ghost $m_t^2\sim\Lambda_3^6/m^4$ does not come close to the cutoff. As already noted, the potentially worrisome terms are the ones involving the highest powers in external momenta.\footnote{Note that from an EFT point of view, even in the present context of loop corrections to the two-point function these higher order terms should be treated as additional vertecies of the theory rather than including them in the propagator.} For instance, a counterterm induced by the $\sim k^8$ contribution after taking the decoupling limit towards the relevant scale behaves like
\be\label{DimExpectk8DL}
\sim\frac{\pd^8}{\Lambda_2^4m^2}\,A^2\xrightarrow[]{\text{DL}}\frac{\pd^6}{\Lambda_3^6}\frac{\pd^2}{m^2}\left(\pd\phi\right)^2\,.
\ee
Hence, in the decoupling limit this term technically blows up and the quantum correction is out of control compared to the classical kinetic term. In the following we will show by explicit calculation that despite this bad expectation based on dimensional analysis, the one-loop corrections precisely organize themselves in such a way that their heavy suppression by powers of the single cutoff scale $\Lambda_3$ is not spoiled by excessive mass terms in the denominator.

In order to evaluate the contributions explicitly one has to perform the usual sum over all possible Wick-contractions which a priory gives $3!^2/2=18$ and $4!/2=12$ possible contractions for each diagram respectively without counting the vertex exchange factor which as usual cancels the prefactor of the exponential expansion. Considering all possible combinations of vertices, summing up all diagrams and following a standard dimensional regularization procedure with $d=4+2\epsilon$\footnote{Note that at one loop the divergent part is blind to the extra factors of $d$ in the Levi-Civita contractions, such that we will disregard them.}, the divergent part of the reduced matrix element up to two powers of momenta reads\footnote{Note the different relative factors compared to equation $(3.13)$ in \cite{Charmchi:2015ggf}. In particular, when trying to reproduce the results of \cite{Charmchi:2015ggf} we already obtain discrepancies in earlier steps, for instance eq. (3.9). Given that our computation \eqref{2ptFeyn} is confirmed by an entirely independent method \eqref{finalResultsG2} we are very confident about our results.}
\begin{IEEEeqnarray}{rCl}\label{2ptFeyn}
\mathcal{M}_2^{\text{div}}&=\frac{\epsilon^{\alpha}_{\scriptstyle k}\,\epsilon^{\beta}_{\scriptstyle -k}}{16\pi^2\epsilon\,\Lambda_2^4}&\left[\,k^2 \eta_{\alpha\beta}\,m^4\left(-3\,c^2_3+6\,\tilde{c}_4-4\,c_3\tilde{c}_5 +2\,\tilde{c}^2_5\right)+ \eta_{\alpha\beta}\,m^6\left(-3\,c^2_3+6\,\tilde{c}_4\right)\right.\nonumber\\
&&\left.\,
+ k_\alpha k_\beta\,m^4\left(12\,c^2_3-6\,\tilde{c}_4+16\,c_3\tilde{c}_5 +\tfrac{11}{2}\,\tilde{c}^2_5\right)\right. \nonumber\\
&&\left.\,
+k^2 k_\alpha k_\beta\,m^2\left(-3\,c_3^2-2\,c_3\tilde{c}_5+\tfrac{2}{3}\,\tilde{c}_5^2\right)-k^4\eta_{\alpha\beta}\,\frac{m^2}{\Lambda_2^4}\,\frac{19}{6}\,\tilde{c}_5^2\right. \nonumber \\
&&\left.\,
+k^4 k_\alpha k_\beta \left(\tfrac{1}{2}\,c_3^2-\tfrac{13}{12}\,\tilde{c}_5^2\right)+k^6\eta_{\alpha\beta}\,\frac{4}{3}\,\tilde{c}_5^2\right. \nonumber \\
&&\left.\,
+k^6\,\frac{1}{m^2}\,\frac{1}{6}\,\tilde{c}_5^2\left(k_\alpha k_\beta-k^2\eta_{\alpha\beta}\right)
 \right] 
\end{IEEEeqnarray}
where $k_1=-k_2=k$.

Observe that $\L_6$ and the term in $\L_4$ proportional to $c_4$ do not contribute at all. Moreover, the counterterm induced by $\tilde{c}_4$ preserves the ghost free structure of the kinetic and mass term $\left(\Box+m^2\right)\,\eta_{\alpha\beta}-\pd_\alpha \pd_\beta$ as could have been expected by the structure of the operator, while the contributions from $\L_3$ and $\L_5$ introduce the anticipated detuning which leads to an operator \eqref{Odetuning}. However, as discussed above, only the terms involving a power of external momenta equal or higher than eight are troublesome. Let's thus focus on the last line in \eqref{2ptFeyn}: The terms with momentas to a power of ten are absent, even though they technically would have been allowed. Hence, the structure of the generalized Proca model is precisely such, that these dangerous corrections are canceled. However, there is still a contribution $\sim k^8$. But remarkably, the theory only allows for this contribution to induce a counterterm with the specific gauge preserving combination $\Box\eta_{\alpha\beta}-\pd_\alpha \pd_\beta$. This cures the EFT structure as can be seen in the decoupling limit, where compared to  \eqref{DimExpectk8DL} we now have
\be\label{DimExpectk6F2DL}
\sim\frac{\pd^6}{\Lambda_2^4m^2}\,F^2\xrightarrow[]{\text{DL}}\frac{\pd^6}{\Lambda_3^6}F^2\,,
\ee
perfectly fitting into the hierarchy between classical and quantum terms.

At this point, the cancellations observed above magically seem to rescue the EFT. In order to better understand these nice properties of the EFT we will change gears in the next section \ref{QuantumStability} and perform a thorough decoupling limit analysis which will allow us to extrapolate quantum stability of the generalized Proca theory in its full generality.

But first, lets also explicitly calculate the higher point one-loop contributions.

\subsubsection{Three-point}
\label{ThreePoint}

\begin{figure}[H]
\begin{center}
\begin{fmffile}{loops3pf}
\begin{fmfgraph*}(125,90)
     \fmfleft{i1,i2}
     \fmfright{o}
     \fmf{plain,tension=3}{i1,v1}
     \fmf{plain,tension=3}{i2,v2}
     \fmf{plain}{v1,v2}
     \fmf{plain}{v1,v3}
     \fmf{plain}{v2,v3}
     \fmf{plain,tension=3}{v3,o}
     \fmfdot{v1,v2,v3}
    \end{fmfgraph*}
    \qquad \qquad
  \begin{fmfgraph*}(125,90)
 	\fmfleft{i1,i2}
     \fmfright{o}
      \fmf{plain,tension=3}{i1,v1}
     \fmf{plain,tension=3}{i2,v1}
     \fmf{plain,left=1}{v1,v2}
     \fmf{plain,left=1}{v2,v1}
     \fmf{plain,tension=3}{v2,o}
     \fmfdot{v1,v2}
\end{fmfgraph*}
\end{fmffile}
\end{center}
\caption{\small{Two distinct one-loop $1$PI diagrams giving rise to corrections of the three point function. The diagrams represents contributions from the four possible combinations out of $\L_{3}$ and $\L_{5}$ and contributions from the mixing of even and odd numbered interaction terms respectively.}}
\label{loop3pf} 
\end{figure}
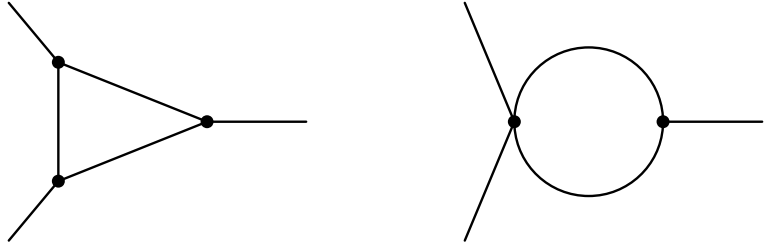

With three external legs there exist as well two distinct $1$PI one-loop diagram structures represented in Fig.\ref{loop3pf}. The first diagram receives contributions from combinations out of $\L_{3}$ and $\L_{5}$ and the second one pairs $\L_{4,6}$ with $\L_{3,5}$.

Again, we can have a look at what awaits us by invoking dimensional analysis together with Lorentz invariance. Note that there are now three indices of external polarization vectors to be contracted.
\small
\begin{equation}\label{3ptDim}
\begin{split} 
\mathcal{M}_{\scriptstyle\L_3^3}&\sim\,\frac{m^6}{\Lambda_2^6}\left(k+\frac{k^3}{m^2}+\frac{k^5}{m^4}+\frac{k^7}{m^6}\right)\,, \\
\mathcal{M}_{\scriptstyle\L_3^2\L_5,\,\L_3\L_4}&\sim\,\frac{m^4}{\Lambda_2^6}\left(m^2k+k^3+\frac{k^5}{m^2}+\frac{k^7}{m^4}+\frac{k^9}{m^6}\right)\,, \\
\mathcal{M}_{\scriptstyle\L_3\L_5^2,\,\L_3\L_6,\,\L_5\L_4}&\sim\,\frac{m^2}{\Lambda_2^6}\left(m^4k+m^3\pd^3+k^5+\frac{k^7}{m^2}+\frac{k^9}{m^4}+\frac{k^{11}}{m^6}\right)\,, \\
\mathcal{M}_{\scriptstyle\L_5^3,\,\L_5\L_6}&\sim\,\frac{1}{\Lambda_2^6}\left(m^6k+m^4k^3+m^3k^5+k^7+\frac{k^9}{m^2}+\frac{k^{11}}{m^4}+\frac{k^{13}}{m^6}\right)\,,
\end{split}
\end{equation}
\normalsize
where again $k$ denote external momenta and the series stops at $1/m^4$ or $1/m^6$ depending on how many propagators are involved. Hence, by the same arguments as above, we should give special attention to the $\L_3\L_5^2$, $\L_5^3$ and $\L_5\L_6$ contributions containing external momenta to the power 11 or higher, as they potentially destabilize the EFT structure.

In order to calculate the diagrams explicitly, let's quickly go through the combinatorics. The first diagram gets four different contributions from combinations out of $\L_{3}$ and $\L_{5}$. For each of these, there are $3!$ possible ways of exchanging the vertecies, which for the $\L_{3}^3$ contribution for example simply cancels the prefactor of the expansion of the exponential. But for the combinations $\L_{3}^2\L_5$ and $\L_{3}\L_5^2$ a redistribution of vertecies leads to three distinct results, hence each of these come only with a vertex exchange factor $2!$. After that, there remains $3!^3$ possible Wick-contractions for each diagram, as the symmetry factor is 1. Note that not all of these contractions are independent of course. 

For the second diagram, there are three distinct channels which need to be considered. For each of these channels at fixed vertecies, there are a priori $72$ different ways of contracting in the S-matrix expansion \eqref{Smatrix} or in other words $3!4!$ different ways of distributing the insertions over the legs divided by the symmetry factor of two. Note that for vertecies with a different number of legs there is no additional vertex exchange factor which could cancel the $1/2!$ in the exponential expansion in \eqref{Smatrix}.

Due to a fast growing complexity of the off-shell expressions, we will only explicitly show here the lowest order momentum result and leave the remaining part in a schematic sum of contributions $M^{\scaleto{(i)\mathstrut}{6pt}}_{3}(\scaleto{c_j\mathstrut}{7pt})$ where $i$ denotes the power of external momenta involved, while the arguments ($c_j$) indicate which diagrams contribute at the given order. In order not to clutter the schematic expansion we will further leave the argument in a general form whenever all possible contributions are involved. The detailed expressions can be found in an ancillary file.
\begin{IEEEeqnarray}{rCl}\label{3ptFeyn}
\mathcal{M}_3^{\text{div}}&=\frac{m^6}{16\pi^2\epsilon\,\Lambda_2^6}&\left[M^{\scaleto{(1)\mathstrut}{6pt}}_{3}(\scaleto{c_3^3,c_3^2\tilde{c}_5,c_3(c_4+\tilde{c}_4),\tilde{c}_5\tilde{c}_4\mathstrut}{9pt})+\frac{1}{m^2}M^{\scaleto{(3)\mathstrut}{6pt}}_{3}(\scaleto{c_j\mathstrut}{9pt})+\frac{1}{m^4}M^{\scaleto{(5)\mathstrut}{6pt}}_{3}(\scaleto{c_j\mathstrut}{9pt})+\frac{1}{m^6}M^{\scaleto{(7)\mathstrut}{6pt}}_{3}(\scaleto{c_j\mathstrut}{9pt})\right.\nonumber\\
&&\left.\,
+\frac{1}{m^8}M^{\scaleto{(9)\mathstrut}{6pt}}_{3}(\scaleto{c_3\tilde{c}_5^2,\tilde{c}_5^3,c_3\tilde{c}_6,\tilde{c}_5\tilde{c}_6\mathstrut}{9pt})\right]\,.
\end{IEEEeqnarray}
with the explicit leading order expression
\begin{IEEEeqnarray}{rCl}
M^{\scaleto{(1)\mathstrut}{6pt}}_{3}=&-2i \bigl(2 c_{3}{}^3 + c_{3}{} (c_{4}{} - 2 \tilde{c}_{4}{}) + 9 c_{3}{}^2 \tilde{c}_{5}{} - 9  \tilde{c}_{5}{}\tilde{c}_{4}{}\bigr) \bigl(\epsilon_{23}{} \epsilon k_{11}{} + \epsilon_{13}{} \epsilon k_{22}{} -  \epsilon_{12}{} (\epsilon k_{31}{} + \epsilon k_{32}{})\bigr)\nonumber\,,
\end{IEEEeqnarray}
where we denote $\epsilon_{ij}\equiv\epsilon_{k_i}\cdot\epsilon_{k_j}$ and $ \epsilon k_{ij}\equiv \epsilon_{k_i}\cdot k_j$.

Regardless of the precise form of the contributions, the important result from \eqref{3ptFeyn} is really the simple absence of any contribution going like $\sim k^{13}$ or $\sim k^{11}$ which would destabilize the EFT.\footnote{Note that with an odd number of external fields we do not have any marginal contribution which should preserve gauge invariance.} Hence, again even though dangerous contributions would technically be allowed \eqref{3ptDim} the calculated series stops at a healthy order.

At this point, one could conclude that all quantum corrections which renormalize the given classical structure involving gauge breaking operators, although being heavily suppressed, come from diagrams involving either $\L_3$ of $\L_5$. In other words, upon a restriction of the generalized Proca model to the even numbered terms $\L_2$, $\L_4$ and $\L_6$ by choosing $c_3=\tilde{c}_5=0$, the only one-loop correction so far is a gauge preserving operator proportional to $\tilde{c}_4$. Moreover, this choice is technically natural, since with only $\L_{4,6}$ insertions no diagrams with an odd number of external legs can be constructed. However, these properties are lost as soon as corrections to higher point functions are taken into account as it will become clear through the 4-point example below.

\subsubsection{Four-point}
\label{FourPoint}

\begin{figure}[H]
\begin{center}
\begin{fmffile}{loops4pf}
\begin{fmfgraph*}(125,50)
     \fmfleft{i1,i2}
     \fmfright{o1,o2}
     \fmf{plain,tension=3}{i1,v1}
      \fmf{plain,tension=3}{i2,v1}
     \fmf{plain,left=1}{v1,v2}
     \fmf{plain,left=1}{v2,v1}
     \fmf{plain,tension=3}{v2,o1}
     \fmf{plain,tension=3}{v2,o2}
     \fmfdot{v1,v2}
    \end{fmfgraph*}
\end{fmffile}
\end{center}
\caption{\small{A $1$PI diagram giving rise to corrections of the four point function. The diagrams represents contributions from the three possible combinations out of $\L_{4}$ and $\L_{6}$.}}
\label{4ptloopdia} 
\end{figure}
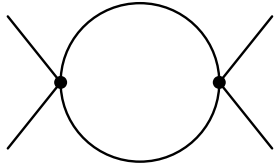

We thus also calculate corrections to the four point function, but restrict ourselves by simplicity to the contributions of the diagram in Fig. \ref{4ptloopdia}. The symmetry factor of the diagram is two, such that for each of the three distinct channels there are at first sight $4!4!/2=288$ possible Wick contractions. Again vertex exchange cancels the $2!$ in the exponential expansion.
\noindent
After adding all the contractions together and going through the dimensional regularization procedure, the schematic form of the result reads
\begin{IEEEeqnarray}{rCl}\label{4ptFeyn}
\mathcal{M}_4^{\text{div}}&=\frac{m^8}{16\pi^2\epsilon\,\Lambda_2^8}&\left[M^{\scaleto{(0)\mathstrut}{6pt}}_{4}(\scaleto{\tilde{c}_4^2\mathstrut}{9pt})+\frac{1}{m^2}M^{\scaleto{(2)\mathstrut}{6pt}}_{4}(\scaleto{(c_4+\tilde{c}_4)^2,\tilde{c}_4\tilde{c}_6\mathstrut}{9pt})+\frac{1}{m^4}M^{\scaleto{(4)\mathstrut}{6pt}}_{4}(\scaleto{c_j^2\mathstrut}{9pt})+\frac{1}{m^6}M^{\scaleto{(6)\mathstrut}{6pt}}_{4}(\scaleto{\tilde{c}_j^2\mathstrut}{9pt})+\frac{1}{m^8}M^{\scaleto{(8)\mathstrut}{6pt}}_{4}(\scaleto{\tilde{c}_j^2\mathstrut}{9pt})\right.\nonumber \\
&&\left.\,
+\frac{1}{m^{10}}M^{\scaleto{(10)\mathstrut}{6pt}}_{4}(\scaleto{(c_4+\tilde{c}_4)\tilde{c}_6,\tilde{c}_6^2\mathstrut}{9pt})+\frac{1}{m^{12}}M^{\scaleto{(12)\mathstrut}{6pt}}_{4}(\scaleto{\tilde{c}_6^2\mathstrut}{9pt})\right]\,,
\end{IEEEeqnarray}
with the same notation as above and details in the ancillary file. Thus also the classical terms $\L_4$ and $\L_6$ get renormalized. In particular, $M^{\scaleto{(2)\mathstrut}{6pt}}_{4}$ generates operators of the form\footnote{As can be conveniently seen in the position space result \eqref{finalResultsG4} of the next subsection.} $A^\mu A^\nu\left(\pd_\mu A_\alpha \pd_\nu A^\alpha+\pd_\alpha A_\mu \pd^\alpha A_\nu\right)$ which destroy the classical ghost-free tuning. Yet again these operators come with heavy suppressions such that the associated ghost degree of freedom will have a mass way above the cutoff. Only the contribution involving twelve external momenta is potentially worrisome as it naively diverges in the decoupling limit. Yet, again, we expect the corresponding counterterm to preserves gauge invariance\footnote{Due to the immense complexity of the expression resulting from the Feynman diagram calculation at this order, there is no use in trying to explicitly show this statement in the present unitary gauge calculation.} such that the actual decoupling limit is of the form
\be\label{DimExpectk6F2DL}
\sim\frac{\pd^8}{\Lambda_2^8m^4}\,F^4\xrightarrow[]{\text{DL}}\frac{\pd^8}{\Lambda_3^8}\frac{F^2}{\Lambda_3^4}\,F^2\,.
\ee
as will become clear in section \ref{QuantumStability} when performing the decoupling limit analysis.

\subsection{Cross-check}
\label{Schw}

As a complementary check, we compute one-loop counterterms using an alternative, effective action based method which combines background field and generalized Schwinger-DeWitt techniques  \cite{BARVINSKY19851}. This method has the additional advantage that it naturally generalizes to curved space-time. The details of the calculation can be found in the appendix \ref{SchwTech}.

Here, we directly skip to the results which serve as highly non trivial checks of the Feynman diagram based momentum space calculations above, as the only common ground of the two methods is the input of the Lagrangian. Due to exceeding computational cost for results at high orders in derivatives, we restricted ourselves to the computation of terms involving a maximum of four derivatives acting on the background fields, which translates into a limitation to four powers of external momentas. However, since Feynman diagram calculations are not structured in an expansion of external momenta\footnote{Meaning that the integration of a given Feynman loop-diagram expression directly gives the result to all orders of external momenta.} the matching of the results at low powers of momenta gives very strong support for the validity of the entire momentum-space calculation. This provides us with confidence for the correctness of our results, especially in comparison with the previous computations in \cite{Charmchi:2015ggf}.

For the full logarithmically divergent one-loop contribution to the 2-point effective action up to a power of four derivatives we find
\begin{IEEEeqnarray}{rCl}\label{finalResultsG2}
\Gamma_{1,2}^{\rm div} &=\frac{m^4}{16 \pi^2\epsilon\,\Lambda_2^4}\,\int \mathrm{d}^4x &\, \left[\;\left(-\frac{3}{2}c_3^2+3\,\tilde{c}_4-2\,c_3\tilde{c}_5+\tilde{c}_5^2\right)\pd_\mu \bar{A}_\nu\pd^\mu \bar{A}^\nu\right.\nonumber\\
&&\left.
+\left(-\frac{3}{2}c_3^2+3\,\tilde{c}_4\,\right)m^2\bar{A}_\mu\bar{A}^\mu\phantom{\tfrac{\tilde{c}_5^2}{\Lambda^4}}\right.\nonumber\\
&&\left.
+\left(\;\;\,6\,c_3^2-3\,\tilde{c}_4+8\,c_3\tilde{c}_5+\frac{11}{4}\tilde{c}_5^2\right)(\pd_\mu \bar{A}^\mu)^2 \right.\nonumber\\
&&\left. 
-\left(\;\;\,\frac{3}{2}c_3^2+c_3\tilde{c}_5-\frac{1}{3}\tilde{c}_5^2\right)\frac{1}{m^2}\pd_\mu\pd_\nu \bar{A}^\nu\Box\bar{A}^\mu\right.\nonumber\\
&&\left. 
-\;\;\;\,\frac{19}{12}\, \tilde{c}_5^2\,\frac{1}{m^2}\Box \bar{A}_\mu\Box\bar{A}^\mu \right].
\end{IEEEeqnarray}
Additionally, we present here a selection of the most relevant 3 and 4-point leading order results up to three powers of external momenta
\footnotesize
\begin{IEEEeqnarray}{rCl}\label{finalResultsH}
\Gamma_{1,3}^{\rm div} &\supset\frac{1}{16 \pi^2\epsilon}\frac{m^6}{\Lambda_2^6}\int \mathrm{d}^4x \,& \bigg[9\,\tilde{c}_4\tilde{c}_5\,\bar{A}^2\pd_\alpha\bar{A}^\alpha+\frac{1}{12}\tilde{c}_5^3\frac{1}{m^2}\Big\{11\left(\pd_\alpha\bar{A}^\alpha\right)^3+51\,\pd_\alpha\bar{A}^\alpha\left(\pd_\mu\bar{A}_\nu\pd^\mu\bar{A}^\nu+\pd_\mu\bar{A}_\nu\pd^\nu\bar{A}^\mu\right)\nonumber\\
&&
-6\,\pd^\mu\bar{A}^\nu\pd_\alpha\bar{A}_\mu\pd^\alpha\bar{A}_\nu-2\,\pd^\mu\bar{A}^\nu\pd_\nu\bar{A}_\alpha\pd^\alpha\bar{A}_\mu\Big\} \bigg]
\label{finalResultsG3}\\
\Gamma_{1,4}^{\rm div} &\supset\frac{1}{16 \pi^2\epsilon}\frac{m^6}{\Lambda_2^8}\int \mathrm{d}^4x \,&\bigg[9\,\tilde{c}_4^2\,m^2\big(\bar{A}^2\big)^2-\left(2\,c_4^2+16\,c_4\tilde{c}_4+20\,\tilde{c}_4^2+3\,\tilde{c}_4\tilde{c}_6\right)\bar{A}^2\, \pd_\alpha\bar{A}_\beta\pd^\beta\bar{A}^\alpha\nonumber\\
&&
-\left(2c_4^2-10c_4\tilde{c}_4-2\tilde{c}_4^2-3\tilde{c}_4\tilde{c}_6\right)\bar{A}^2\left(\pd_\alpha\bar{A}^\alpha\right)^2+2\left(c_4^2+3c_4\tilde{c}_4+9\tilde{c}_4^2\right)\bar{A}^2\,\pd_\alpha\bar{A}_\beta\pd^\alpha\bar{A}^\beta\nonumber\\
&&
-2\left(c_4^2+3c_4\tilde{c}_4+3\tilde{c}_4^2\right)\bar{A}^\mu\bar{A}^\nu\pd_\mu\bar{A}_\alpha\pd_\nu\bar{A}^\alpha+4\left(4c_4^2+\tilde{c}_4^2\right)\bar{A}^\mu\bar{A}^\nu\pd_\mu\bar{A}_\nu\pd_\alpha\bar{A}^\alpha\nonumber\\
&&
-2\left(2c_4^2-3c_4\tilde{c}_4+5\tilde{c}_4^2\right)\bar{A}^\mu\bar{A}^\nu\pd_\alpha\bar{A}_\mu\pd^\alpha\bar{A}_\nu\bigg].\label{finalResultsG4}
\end{IEEEeqnarray}
\normalsize

In order to relate these results to the Feynman diagram calculations in \S\ref{Feyn} recall that the effective action is a generating functional of $1$PI correlation functions
\be
\frac{\delta^n\Gamma[\bar{\pi}]}{\delta\bar{A}^{\mu_1}(x_1)...\delta\bar{A}^{\mu_n}(x_n)}\biggr\rvert_{\bar{A}=\langle A\rangle} =\langle A_{\mu_1}(x_1)...A_{\mu_n}(x_n) \rangle_{1\rm PI}\,.
\ee
The $1$PI correlation functions in turn are given by the sum of all $1$PI diagrams with $n$ external points. Thus, fourier transformed functional derivatives of divergent one-loop effective action results at vanishing mean field should coincide with the corresponding divergent off-shell results of the $1$PI diagrams calculated in \S\ref{Feyn}. We explicitly checked this for all the expressions above. For instance, for the 2-point result \eqref{finalResultsG2} it can be seen by eye that it precisely matches the momentum space calculation \eqref{2ptFeyn} as the conversion merely introduces a factor of $1/2$. 


\section{Decoupling Limit Analysis}
\label{QuantumStability}

Starting from the above one-loop results we will now intent a complete radiative stability analysis of the Generalized Proca EFT. To this end we will leave the unitary gauge employed in the previous section and instead take the decoupling limit already at the level of the Lagrangian. In a first step, this will allow us in \S\ref{reint} to shed light on the observed cancellations in the unitary gauge calculations by reobtaining the most important aspects of the results in \S\ref{OneLoop} still within our restricted generalized Proca model \eqref{FunctionChoice}. At the same time, this confirms that just as within massive gravity theories (see eg. \cite{Hinterbichler:2011tt,deRham:2013qqa}) taking the decoupling limit and computing quantum corrections are two operations which commute. Based on this knowledge, we will in a second step \S\ref{fullQS} establish healthiness of the full generalized Proca theory \eqref{Lagrangians} under quantum corrections at all orders by showing that all these models admit a well defined decoupling limit where classical and quantum operators are structured in a well defined hierarchy. In particular we will identify the correct classical and quantum expansion parameters of the theory. It turns out that the structure of the generalized Proca EFT bares many similarities with equally self-interacting non-abelian $SU(2)$ spin $1$ fields endowed with a mass term and thus the hierarchy structure of the weak sector of the Standard Model EFT.

\subsection{Reinterpretation of the Unitary Gauge Results}
\label{reint}

As discussed in \S\ref{GPm}, rewriting the generalized Proca model \eqref{action} by introducing a St\"uckelberg field $\phi$ allows one to take a smooth $m\rightarrow 0$ limit, without loosing any degrees of freedom. In the present interacting theory, the decoupling limit works actually as a high energy limit way above the vector mass and right at the lowest cutoff $\Lambda_3\equiv(\Lambda_2^2m)^{\frac{1}{3}}$. The only operators which survive this high energy limit are the least suppressed ones and thus the decoupling limit puts focus on the operators with the poorest behavior. The resulting theory is given by \eqref{actionDL}, where the transverse modes, in this limit described by a massless and gauge invariant vector field $A_\mu$, are decoupled in a symmetry sense from the longitudinal Goldstone mode $\phi$. 

In this section we will show how to infer the explicitly computed general structure of highest order operators in the unitary gauge directly from the decoupling limit. This means that in general, from the healthiness of the EFT in the decoupling limit we can predict the qualitative aspect of the cancellations of leading order terms in the unitary gauge observed in the previous section \ref{OneLoop}.

But first of all, as $\L_3$ and part of $\L_4$ reduce to scalar Galileon terms in the decoupling limit, it is useful to quickly remind ourselves how the Galileon EFT is structured. It is well known that the classical terms are not renormalized, since quantum corrections always come with more derivatives per field. Employing dimensional regularization the full scalar Galileon EFT lagrangian schematically goes like \cite{Nicolis:2004qq}
\be\label{Fgal}
\L^{\text{r}}_{\text{Gal}}\sim(\partial \phi)^2\left(\frac{\partial^2\phi}{\Lambda_3^3}\right)^i + \left(\frac{\partial^{2}}{\Lambda_3^{2}}\right)^{3+n}(\partial \phi)^2\left(\frac{\partial^2\phi}{\Lambda_3^3}\right)^{m} \,, \quad 3\geq i\geq0\,,\;n,m\geq0\,,
\ee
where the first set of terms are the classical operators of the theory which induce the second set of operators through quantum loops at all orders. This essentially follows from Lorentz invariance, the high energy behavior of the massless propagator and the fact that only the log divergent piece enters in the construction of counterterms. The full Lagrangian written as \eqref{Fgal} defines the two expansion parameters
\be\label{GalParameters}
\alpha_{\text{cl}}=\frac{\partial^2\phi}{\Lambda_3^3}\quad \text{and}\quad \alpha_{\text{q}}=\frac{\partial^2}{\Lambda_3^2}\,,
\ee
which allow to clearly distinguish between classical and quantum operators and thus assures radiative stability. In other words, there exists a regime below the UV cutoff, where the a priori irrelevant classical non-linear galileon operators become important compared to the kinetic term $\alpha_{\text{cl}}\sim\mathcal{O}(1)$, while quantum corrections are still under control $\alpha_{\text{q}} \ll 1$. 

Coming back to the generalized Proca theory in the decoupling limit, the considerations above directly imply that in the full theory, for instance also in the unitary gauge, all quantum corrections generated exclusively through $\L_3$ and the $c_4$ term in $\L_4$ are safe. Translating back decoupling limit results to the unitary gauge is schematically done via a replacement $\pd\phi\rightarrow m\,A$, such that from the second set of terms in \eqref{Fgal} the corresponding highest order quantum correction in the unitary gauge are
\be\label{FL3L4}
\L^{\text{c}}_{\L_3,\L_4(c_4)}\sim m^m\frac{\pd^{4+2n+m}}{\Lambda_3^{2n+3m}}\,A^m \,, \quad n\geq0\,,\;m\geq2\,,
\ee
where for example the case $m=2$, $n=0$ corresponds to the two-point result \S\ref{TwoPoint}. Moreover, with this knowledge we could also have directly inferred the absence of high momenta power contributions proportional to $c_4$ in \eqref{2ptFeyn}, because $\L_{4\,\text{Gal}}$ does not contribute to the one-loop correction with two exernal legs.

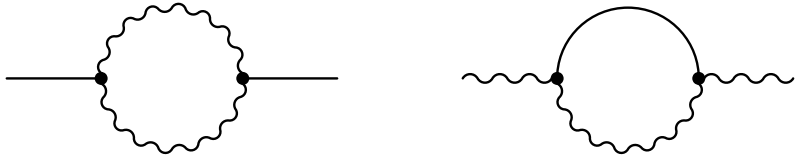
\begin{figure}[H]
\begin{center}
\begin{fmffile}{loops2pfDL}
\begin{fmfgraph*}(125,50)
     \fmfleft{i}
     \fmfright{o}
     \fmf{plain,tension=3}{i,v1}
     \fmf{photon,left=1}{v1,v2}
     \fmf{photon,left=1}{v2,v1}
     \fmf{plain,tension=3}{v2,o}
     \fmfdot{v1,v2}
    \end{fmfgraph*}
    \qquad \qquad
  \begin{fmfgraph*}(125,50)
     \fmfleft{i}
     \fmfright{o}
     \fmf{photon,tension=3}{i,v1}
     \fmf{plain,left=1}{v1,v2}
     \fmf{photon,left=1}{v2,v1}
     \fmf{photon,tension=3}{v2,o}
     \fmfdot{v1,v2}
    \end{fmfgraph*}
\end{fmffile}
\end{center}
\caption{\small{Two distinct one-loop $\L_{5}$ diagram contributions in the decoupling limit giving rise to corrections of the two point function. Solid lines represent scalar legs or propagators. Each external leg comes with two derivatives applied on the field $\sim\pd^2\phi$. Wiggled lines correspond to gauge preserving vector legs $\sim F$ or corresponding propagators. In the decoupling limit, propagators have a good $\sim 1/p^2$ high energy behavior.}}
\label{2ptloopdiaDL} 
\end{figure}

Up to now, these considerations parallel the powercounting arguments in the unitary gauge. However, recall that quantum corrections involving $\L_5$ and $\L_6$ required non-trivial cancellations of the leading order estimates in order to remain healthy. These cancellations can readily be explained from the point of view of the decoupling limit. Let's first focus at the one loop corrections to the propagator. $\L_5$ in the decoupling limit is an interaction term between the massless vector and the Goldstone \eqref{actionDL}. This directly implies that in the high energy limit no one-loop diagram can be formed between the two terms $\L_3\L_5$ and hence, terms proportional to $c_3\tilde{c}_5$ have no impact close to the cutoff scale $\Lambda_3$ and remain highly suppressed. This is in perfect agreement with the obtained results \eqref{2ptFeyn}. However, two distinct diagrams can be formed with two $\L_5$ insertions depicted in Fig.\ref{2ptloopdiaDL}, where straight lines denote scalar legs $\sim\pd^2\phi$ or scalar propagators $\sim 1/p^2$ and the wiggled lines massless vector legs and propagators $\sim F$ and $\sim 1/p^2$ respectively.

\noindent
The first diagram induces a counterterm proportional to $(\pd^2\phi)^2$ since just as in the pure galileon case the external legs carry two derivatives per scalar field. This is a contribution of the same order as \eqref{Fgal}. On the other hand, because of the two vector legs, the contribution from second diagram is bound to give rise to gauge invariant operator, such that in total we have
\be\label{ctDL1}
\mathcal{L}^{\text{c DL}}_{\L_5\L_5}\sim \tilde{c}_5^2\left[\frac{\pd^4}{\Lambda_3^6}\left(\pd^2\phi\right)^2+\frac{\pd^6}{\Lambda_3^6}\left(F\right)^2\right]\,.
\ee
We can now understand why there could not be a contribution proportional to $\sim k^{10}$ in the one-loop two-point corrections in the original theory \eqref{2ptFeyn} simply from the fact that terms of this order cannot be formed in the decoupling limit. Again translating back to the unitary gauge through $\pd\phi\rightarrow m\,A$ leads to
\be
\L^{\text{c}}_{\L_5\L_5}\sim \tilde{c}_5^2\left[\frac{m^2\pd^4}{\Lambda_3^6}\left(\pd A\right)^2+\frac{\pd^6}{\Lambda_3^6}\left(F\right)^2\right]\,,
\ee
which is qualitatively in perfect agreement with the $F^2$ structure obtained in the last line of \eqref{2ptFeyn}.

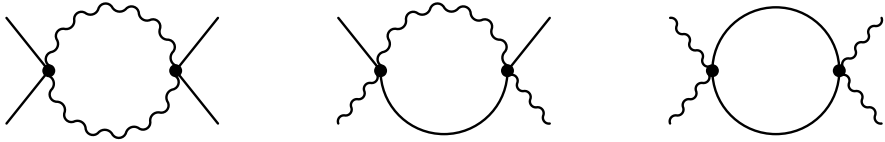
\begin{figure}[H]
\begin{center}
\begin{fmffile}{loops4pfDL}
\begin{fmfgraph*}(100,40)
     \fmfleft{i1,i2}
     \fmfright{o1,o2}
     \fmf{plain,tension=3}{i1,v1}
      \fmf{plain,tension=3}{i2,v1}
     \fmf{photon,left=1}{v1,v2}
     \fmf{photon,left=1}{v2,v1}
     \fmf{plain,tension=3}{v2,o1}
     \fmf{plain,tension=3}{v2,o2}
     \fmfdot{v1,v2}
    \end{fmfgraph*}
    \qquad
  \begin{fmfgraph*}(100,40)
     \fmfleft{i1,i2}
     \fmfright{o1,o2}
     \fmf{photon,tension=3}{i1,v1}
      \fmf{plain,tension=3}{i2,v1}
     \fmf{photon,left=1}{v1,v2}
     \fmf{plain,left=1}{v2,v1}
     \fmf{photon,tension=3}{v2,o1}
     \fmf{plain,tension=3}{v2,o2}
     \fmfdot{v1,v2}
    \end{fmfgraph*}
    \qquad
  \begin{fmfgraph*}(100,40)
     \fmfleft{i1,i2}
     \fmfright{o1,o2}
     \fmf{photon,tension=3}{i1,v1}
      \fmf{photon,tension=3}{i2,v1}
     \fmf{plain,left=1}{v1,v2}
     \fmf{plain,left=1}{v2,v1}
     \fmf{photon,tension=3}{v2,o1}
     \fmf{photon,tension=3}{v2,o2}
     \fmfdot{v1,v2}
    \end{fmfgraph*}
\end{fmffile}
\end{center}
\caption{\small{Three distinct one-loop $\L_{6}$ diagram contributions in the decoupling limit giving rise to corrections of the four point function. Solid lines represent scalar legs or propagators. Each external leg comes with two derivatives applied on the external field $\sim\pd^2\phi$. Wiggled lines correspond to gauge preserving vector legs $\sim F$ or corresponding propagators. In the decoupling limit, propagators have a good $\sim 1/p^2$ high energy behavior.}}
\label{4ptloopdiaDL} 
\end{figure}

Similar for higher point functions. For example $\tilde{c}_5^3$ diagrams at most generate counterterms of the form \small$\sim\tfrac{1}{\Lambda_3^9}\pd^6(\pd^2\phi)F^2$\normalsize, while it is not possible to form contributions going like \small$\sim\tfrac{1}{\Lambda_3^9}\pd^8F^3$\normalsize. This is explicitly confirmed by the 3-point calculation \eqref{3ptFeyn}. In a similar manner, the 4-point results can be understood. For instance, the $\tilde{c}_6^2$ high energy contributions come from the three diagrams \ref{4ptloopdiaDL} in the decoupling limit. The schematic form of the corresponding counterterms is
\be\label{ctDL2}
\mathcal{L}^{\text{c DL}}_{\L_6\L_6}\sim \tilde{c}_6^2\left[\frac{\pd^4}{\Lambda_3^{12}}\left(\pd^2\phi\right)^4+\frac{\pd^6}{\Lambda_3^{12}}\left(\pd^2\phi\right)^2(F)^2+\frac{\pd^8}{\Lambda_3^{12}}\left(F\right)^4\right]\,.
\ee
In unitary gauge this corresponds to
\be\label{UnitaryGaugeL5}
\L^{\text{c}}_{\L_6\L_6}\sim \tilde{c}_6^2\left[\frac{m^4\pd^4}{\Lambda_3^{12}}\left(\pd A\right)^4+\frac{m^2\pd^6}{\Lambda_3^{12}}\left(\pd A\right)^2\left(F\right)^2+\frac{\pd^9}{\Lambda_3^{12}}\left(F\right)^4\right]\,,
\ee
which for instance shows that $M^{\scaleto{(12)\mathstrut}{6pt}}_{4}$ in \eqref{4ptFeyn} indeed possesses a gauge invariant structure.

\subsection{Quantum Stability of Generalized Proca Theories}\label{fullQS}

These decoupling limit arguments can be generalized to all higher point functions, indicating that the EFT structure is stable under all possible quantum corrections from an EFT point of view. The following argument will even go beyond the specific model chosen in \eqref{FunctionChoice} and include all possible generalized Proca EFT's. 

First of all note that up to total derivatives, a generic generalized Proca theory with classical action \eqref{Lagrangians} can be expanded in a sum of terms with a schematic form\footnote{We assume here implicitly, that the generic functions $f$ in \eqref{Lagrangians} are smooth, such that they admit a Taylor series expansion.}
\be\label{SchematicL}
\mathcal{L}\sim \left(F^2+m^2 A^2\right)\left(\frac{mA}{\Lambda_2^2}\right)^{2a_1}\left(\frac{F}{\Lambda_2^2}\right)^{a_2}\left(\frac{\pd A}{\Lambda_2^2}\right)^{a_3}\,,\quad a_{1,2,3}\geq0\,,\;a_3\leq 4\, ,
\ee
where of course all suppressed Lorentz indices need to be contracted and the classical structure is such that the theory only propagates the required three degrees of freedom. The first two operators in \eqref{SchematicL} take care of the dimension, while combinations of the dimensionless factors with powers $a_i$ can be divided into three classes of terms:
\begin{itemize}
\item Terms coming from expanding the function $f_2$ in terms of it's arguments, which includes the kinetic and mass term. In \eqref{SchematicL} these terms correspond to operators with $a_3=0$, where the kinetic and mass term are the ones given by $a_1=a_2=a_3=0$. 

\small(On top of the basic $\L_2$ terms, in the specific model \eqref{action} employed in the major part of this work also the term proportional to $\tilde{c}_4$ correspond to this class.)\normalsize
\item Terms resulting from the expansion of the functions $f_i$, $i=3,..,6$. These terms correspond to Galileon like contributions. They do not include any powers of $a_2$ and are proportional to $m^2A^2$ only. 

\small(In \eqref{action} these correspond to the $c_3$ and $c_4$ terms.)\normalsize
\item Terms coming from expanding the functions $\tilde{f}_i$, $i=4,5,6$. These terms correspond to genuinely new vector derivative self interactions with $a_2=0$ as well  but proportional to $F^2$. 

\small(In \eqref{action} these correspond to the terms proportional to $\tilde{c}_5$ and $\tilde{c}_6$.)\normalsize
\end{itemize}
This justifies the specific choice of our model \eqref{FunctionChoice} in retrospective, since we cover all interesting cases. 

The St\"uckelberg trick described in \S\ref{GPm} can now be applied to each classical term in \eqref{SchematicL} which reformulates the theory using gauge redundancy. This alternative formulation of a generic generalized Proca term then allows to perform the decoupling limit \eqref{DL} on each individual operator. We obtain that indeed the limit is well defined and the classical Lagrangian reduces to
\be\label{SchematicLDL}
\mathcal{L}_{\text{DL}}\sim \left(F^2+(\pd\phi)^2\right)\left(\frac{\pd^2\phi}{\Lambda_3^3}\right)^{a_3}\,,\quad 3\geq a_{3}\geq0\,,
\ee
where terms proportional to $F^2$ are limited to $a_{3}\leq 2$. Moreover, all the operators proportional to $f_6(x)$ in \eqref{Lagrangians} are actually total derivatives \cite{Heisenberg:2014rta,Jimenez:2016isa,Heisenberg:2018vsk,Jimenez:2019hpl}, such that in particular also the operator which would lead to a $a_{3}=4$ decoupling limit contribution has vanishing equations of motion. This nicely reflects the fact that for the terms involving only the scalar field $\phi$ the individual classical operators only remain ghost-free in the specific scalar Galileon form \eqref{Fgal}.

Including now loop contributions the specific form of \eqref{SchematicLDL} implies that each vertex in the decoupling limit comes at least with a factor of $1/\Lambda_3^3$. This means that in dimensional regularization and only considering $1$PI diagrams at one loop, where each vertex at least includes one external leg while two legs are contributing to the loop, there are only two distinct schematic building blocks for quantum induced operators \small$\frac{\pd F}{\Lambda_3^3}\;$\normalsize and \small$\frac{\pd^2\phi}{\Lambda_3^3}\;$\normalsize. Therefore, a general one loop counterterm in the decoupling limit has the generic form\footnote{Once again this schematic form is fixed through Lorentz invariance, powercounting and the well behaved propagators in the decoupling limit}
\be\label{fullDL}
\mathcal{L}^{\text{c}}_{\text{DL}}\sim \pd^4\left(\frac{\pd F}{\Lambda_3^{3}}\right)^{2 b_2}\left(\frac{\pd^2\phi}{\Lambda_3^{3}}\right)^{b_3}\sim
\begin{cases}
F^2\left(\frac{\pd^2}{\Lambda_3^2}\right)^{2+b_2}\left(\frac{F^2}{\Lambda_3^4}\right)^{b_2-1}\left(\frac{\pd^2\phi}{\Lambda_3^3}\right)^{b_3}&,\quad b_2\geq 1\\
\\
(\pd\phi)^2\left(\frac{\pd^2}{\Lambda_3^2}\right)^{3+b_2}\left(\frac{F^2}{\Lambda_3^4}\right)^{b_2}\left(\frac{\pd^2\phi}{\Lambda_3^3}\right)^{b_3-2}&,\quad b_3\geq 2
\end{cases} 
\ee
where $2 b_2+b_3=N\geq 2$ with $N$ the number of external fields and $b_{2,3}\geq0$ positive integers.\footnote{The two possible reformulations in \eqref{fullDL} result from comparing the operator to the two kinetic terms $F^2$ and $(\pd\phi)^2$ of the theory. For most $b_i$ values either one can be employed. Only for $b_1=0$ the upper one looses its sense, while the lower one is not valid whenever $b_3<2$.} For example, the cases $\{b_2=0,b_3=2\}$, $\{b_2=1,b_3=0\}$ correspond to the counterterms we already encountered in \eqref{ctDL1}, while $\{b_2=0,b_3=4\}$, $\{b_2=1,b_3=2\}$ and $\{b_2=2,b_3=0\}$ are covered in \eqref{ctDL2}. Thus, on top of the two expansion parameters $\alpha_{\text{cl}}$ and $\alpha_{\text{q}}$ defined in the pure scalar Galileon context \eqref{GalParameters}, we identify a second quantum expansion parameter 
\be
\alpha_{\tilde{\text{q}}}=\frac{F^2}{\Lambda_3^4}\,.
\ee
One can easily generalize this analysis to higher loops. Each additional loop comes with an increase in factors of $1/\Lambda_3^3$ compared to the same diagram without the additional loop. This is because in order to add a loop to a diagram while keeping the number of external legs fixed necessarily requires the inclusion of an additional vertex or the addition of legs to existing vertecies. Now since the number of external legs remains the same in this comparison, to match dimensions these factors can only be compensated with additional powers of derivatives. Thus, higher loops will merely introduce additional factors of $\alpha_{\text{q}}$.\footnote{Actually, Lorentz invariance requires the additional factor to be $\pd^6/\Lambda_3^6$.}

From here on, the analysis exactly parallels the one employed for the consolidation of radiative stability of various derivative self-interacting theories such as scalar Galileons \cite{Luty:2003vm,Nicolis:2004qq,Burgess:2006bm,Hinterbichler:2010xn,Hinterbichler:2011tt,deRham:2012ew,Goon:2016ihr,HorndeskiSurvivals}: The complete EFT Lagrangian can be written as an expansion in the three parameters $\alpha_{\text{cl}}$, $\alpha_{\text{q}}$ and $\alpha_{\tilde{\text{q}}}$ equivalent to the Galileon case \eqref{Fgal}
\be
\mathcal{L}_{\text{DL}}\sim\left(F^2+(\pd\phi)^2\right)\,\alpha_{\text{cl}}^{a_3}+\left(F^2+(\pd\phi)^2\right)\,\alpha_{\text{q}}^{2+n}\alpha_{\tilde{\text{q}}}^l\,\alpha_{\text{cl}}^m\,,\quad 3\geq a_3\geq0\,,\;l,n,m\geq0\,,
\ee
where only the quantum induced operators carry the quantum parameters $\alpha_{\text{q},\tilde{\text{q}}}$. More precisely, every loop operator inevitably carries a non-zero power of these quantum parameters. This marks a clear separation between classical and quantum terms and implies non-renormalization of classical terms.\footnote{Non-renormalization in the weak sense, tied to dimensional regularization (see \cite{Goon:2016ihr}).} Just as in the scalar Galileon case, there exists a regime below the energy scale $\Lambda_3$ where quantum contributions are heavily suppressed $\alpha_{\text{q},\tilde{\text{q}}}\ll 1$, while classical non-linear terms, although equally non-renormalizable, are important compared to the kinetic term $\alpha_{\text{cl}}\sim\mathcal{O}(1)$. Hence, in the decoupling limit, the theory is stable under quantum corrections which directly implies quantum stability of the whole theory as already discussed. Hence, the generalized Proca EFT \eqref{action} does not loose it's key properties when including quantum corrections in their full generality and the effective description is theoretically viable.

Moreover, the commutativity of decoupling limit and quantum correction calculations allows to translate the expansion \eqref{fullDL} back to the unitary gauge, from which one can infer the cancellation of dangerous leading order terms of loop corrections in generic unitary gauge calculations. In particular, this gives access to the least suppressed quantum corrections in the original formulation
\be
\mathcal{L}^{\text{c}}\sim \pd^4\left(\frac{\pd F}{\Lambda_3^{3}}\right)^{2b_2}\left(\frac{m\,\pd A}{\Lambda_3^{3}}\right)^{b_3}\,,
\ee
which are the ones with $b_3=0$, hence the ones which preserve gauge invariance. Other contributions with non-zero $b_3$ and operators which do not survive the decoupling limit are further suppressed by factors of $m/\Lambda_3$.

\section{Conclusion}
\label{discussion}

The search for viable extensions to general relativity is guided on the one hand by observational constraints and the requirement of theoretical consistency on the other. As concerns the latter, the modern understanding of renormalization views in particular gravity theories as effective, such that a 	quantum stability check is indispensable for every proposed model. The sole classical description of an effective field theory does not make sense on a fundamental level. This is especially true for Galileon type models involving irrelevant derivative self-interactions which gain importance only in regimes where loop corrections might harm the classical EFT structure.

In this work we have investigated the stability of generalized Proca theories under quantum corrections and explicitly calculated all one-loop counterterms up to the three-point function with a glimpse towards four-point results. Doing so revealed a generic neutralization of dangerous leading order corrections, preserving the hierarchy between scales and the specific structure of classical operators. These results were confirmed by the use of an entirely independent Heat Kernel method with only the input of the Lagrangian as a common ground. More than a thorough check of the results, this method paves the way to a covariant generalization. Beyond that, a reformulation of the theory in terms of the St\"uckelberg method permitted an interpretation of the results from a different angle heavily relying on the existence and consistency of the decoupling limit in generalized Proca theories. In this limit, one puts the focus on the highest possible scale of the theory, such that radiative stability in the decoupling limit implies radiative stability of the whole EFT. Moreover, the theory looses it's mass scale and the field propagators are modified to have a well behaved $\sim 1/p^2$ high energy behavior. This implies that standard dimensional analysis looses it's subtleties which enabled us to establish a clear hierarchy between classical and quantum operators by the explicit identification of global classical and quantum expansion parameters of the theory. In turn, this empowered us to an extension of the unitary gauge results to arbitrary orders within the decoupling limit, without any restrictions regarding specific model choices. Hence, as an EFT the attractive properties of generalized Proca models withstand the quantum check in its full generality.

These results are especially noteworthy with possible gravitational and cosmological applications in mind, as the hierarchy between classical and quantum non-linearities allow for regimes in which the former dominate while the EFT description is still protected against quantum detuning. Including gravity and matter fields, this endows the theory with a natural Vainshtein screening in dense regions whereas the additional vector field serves as a generalization of gravity on cosmological scales. While we leave the explicit coupling of the full EFT to gravity for future work, we should expect a smooth inclusion of graviton loops as each mixed vertex comes with a heavy plank mass suppression. 


\section*{Acknowledgments}
We would like to thank Jasmin Allenspach and Michael Ruf for useful discussions.
LH is supported by funding from the European Research Council (ERC) under the European Unions Horizon 2020 research and innovation programme grant agreement No 801781 and by the Swiss National Science Foundation grant 179740. 

\newpage


\appendix
\section{Generalized Schwinger-DeWitt Technique} \label{appendix}
\label{SchwTech}

We present here the covariant Schwinger-DeWitt method used in \S\ref{Schw} to cross-check the feynman diagram calculations.\footnote{See \cite{Ruf:2018vzq} for a specific application of the method to a similar theory.} The starting point is the one-loop effective action given by
\be
\label{effA}
\Gamma_{\text{div}}^{(1)}=\frac{i}{2}\Tr \ln \hat{\mathcal{F}}\, ,
\ee
computed after a split of the field $A_\mu\rightarrow\bar{A}_\mu+B_\mu$ into background and quantum parts with $\hat{\mathcal{F}}$ denoting the bilinear form of the action \eqref{action}
\be
\label{split}
S^{(2)}=-\frac{1}{2}\int d^4 x\, B\, \hat{\mathcal{F}}\, B\, , \quad \hat{\mathcal{F}}=\hat{D}_2+\hat{P}\, ,
\ee
 which can be decomposed into it's principle part $[\hat{D}_2]_{\mu\nu}=\mathcal(\Box+m^2)\eta_{\mu\nu}-\pd_\mu\pd_\nu$ and the subleading perturbations $\hat{P}=\sum_{i=3}^6\,\hat{D}_i(\bar{A})$ depending on the background field which originate from the interaction terms $\L_{3,..6}$.

The decomposition \eqref{split} together with an expansion of the logarithm in \eqref{effA} leads to
\be
\label{traceExp}
\Tr \ln \hat{\mathcal{F}}=\Tr \ln \hat{D}_2+\Tr\left[\hat{\mathcal{P}}\hat{D}_2^{-1}\right]-\frac{1}{2}\Tr\left[\hat{P}\hat{D}_2^{-1}\hat{P}\hat{D}_2^{-1}\right]+\mathcal{O}(\hat{P}^3)\, ,
\ee
where the principle operator can be inverted to give
\be
[\hat{D}_2^{-1}]^{\mu\nu}=\frac{1}{\Box+m^2}\left(\eta^{\mu\nu}+\frac{\pd^\mu\pd^\nu}{m^2}\right).
\ee

The trick is now to transform the expansion \eqref{traceExp} above into a sum of terms proportional to universal functional traces whose divergent part can be evaluated by resorting to Schwinger-DeWitt techniques \cite{BARVINSKY19851}. In flat spacetime, the only non-vanishing universal functional traces in dimensional regularization with $d=4-2\epsilon$ are
\small
\be
\label{UFT}
\Tr\;\mathcal{P}^{\mu_{\scaleto{1\mathstrut}{4pt}} ... \mu_{\scaleto{2N\mathstrut}{4pt}}}(\bar{A})\,\pd_{\mu_{\scaleto{1\mathstrut}{4pt}}}...\pd_{\mu_{\scaleto{2N\mathstrut}{4pt}}}\,\frac{1}{(\Box+m^2)^n}\bigg\rvert_{\text{div}}=
\,\frac{i}{16\pi^2\,\epsilon}\,\int\mathrm{d}^{4}x\,\mathcal{P}^{\mu_{\scaleto{1\mathstrut}{4pt}} ... \mu_{\scaleto{2N\mathstrut}{4pt}}}(\bar{A})\,\frac{(-1)^{n}\,m^{2l}}{2^{N}\,l!(n-1)!}\,\eta^{(N)}_{\mu_{\scaleto{1\mathstrut}{4pt}} ... \mu_{\scaleto{2N\mathstrut}{4pt}}}\, ,
\ee
\normalsize
where $2N=2n-4+2l$, $N\geq1$, $n\geq1$, $l=0,1,2,...$ and $\eta^{(n-2+l)}_{\mu_{\scaleto{1\mathstrut}{4pt}} ... \mu_{\scaleto{2n-4+2l\mathstrut}{4pt}}}$ is the totally symmetrized product of $n-2+l$ metrics. Note that the background field dependent piece $\mathcal{P}(\bar{A})$ just goes along the ride, regardless of it's precise structure. 

The terms appearing in the expansion \eqref{traceExp} are cast into the specific form appearing on the left hand side of \eqref{UFT} by commuting all the operators $1/(\Box+m^2)$ to the right. This procedure is efficient, as each commutation decreases the number of partial derivatives in the numerator of \eqref{UFT} compared to the factors of $1/(\Box+m^2)$ and increases the number of derivatives on the background operator $\hat{P}$:
\be\label{Com1}
\left[\frac{1}{\Box+m^2}\,,\hat{P}\right]=-\,\frac{1}{\Box+m^2}\,[\Box\,,\hat{P}]\,\frac{1}{\Box+m^2}\;,\;\text{where}\quad [\Box\,,\hat{P}]=(\Box\hat{P})+2(\pd^\alpha\hat{P})\pd_\alpha
\ee
This means that the while the log expansion \eqref{traceExp} will be cut off by the maximum number of background fields one is interested in, the iterative commutation of operators \eqref{Com1} will constantly increase the number of derivatives applied on the background fields, which thus allows for the computation of counterterms up to any desired but fixed order in derivatives as well as in the fields.

Note that in contrast to the massless case, the expansion in factors of $\frac{m^2}{\Box}$ measured by the integer $l$ allows for divergent contributions of the linear terms in \eqref{traceExp} with $n=1$. However, tadpole contributions arising from the interaction terms $\L_{3}$ and $\L_{5}$ are immediately ruled out by the odd number of derivative factors. Thus, the linear terms will only provide potential corrections to the two point function via contributions from $\L_{4}$ and $\L_{6}$. 
The next terms in the log expansion \eqref{traceExp} $\sim \hat{P}^2$ give rise to contribution to the 2-point function originating in the interactions $\L_{3}$ and $\L_{5}$ and contributions to the 3- and 4-point functions by a suitable mixing of all interaction terms.
As concerns higher point results, we won't need terms in the log expansion \eqref{traceExp} higher than $\sim \hat{P}^3$, as these cover all cases considered through Feynman calculations in \S\ref{Feyn}.

Equation \eqref{finalResultsG2} shows the full log divergent 2-point result up to the given order in derivatives, while a selection of higher point results are given in \eqref{finalResultsH}.


\newpage
\bibliographystyle{utphys}
\bibliography{references} 
\end{document}